\documentclass[journal]{IEEEtran}
\usepackage[T1]{fontenc}

\usepackage{cite}
\ifCLASSINFOpdf
   \usepackage[pdftex]{graphicx}
   \usepackage{epstopdf}
\else
\fi
\usepackage{amsmath}
\usepackage{cleveref}  
\usepackage{array}
\begin{document}
\title{Load Scheduling for Pulse Charging to Flatten Aggregate Power Demand}
\author{Yu~Liu
\thanks{Manuscript submitted on March 31, 2026. ({\it Corresponding author: Yu Liu.})}
\thanks{The author is with the College 
of Urban Transportation and Logistics, Shenzhen Technology University,
Shenzhen, Guangdong, China.}
}

\markboth{Journal of \LaTeX\ Class Files,~Vol.~1, No.~1, May ~2026}%
{Shell \MakeLowercase{\textit{et al.}}: Bare Demo of IEEEtran.cls for IEEE Journals}

\maketitle

\begin{abstract}
Pulse charging can be used to boost up charging speed for lithium-ion batteries and delay battery capacity fading by periodically pausing the current during charging. However, this technique introduces intermittence for current and may thus challenge the electric stability of charger as well as its energy supply source. To deal with this challenge, a coordination method for multiple loads simultaneously being charged has been proposed in this paper. The method exploits the off-time intervals of pulse current to charge other loads. By properly grouping and coordinating the charging loads, the fluctuation and amplitude of the charging current can be mitigated. To optimally schedule all charging loads, mathematical models are formulated to help find out the best scheduling scheme for the loads. Two scenarios have been considered as well as two mathematical models have been proposed and elucidated in the paper. In one scenario all loads are charged using PCs with the same frequency, while in the other scenario PCs with various frequencies are considered. In addition, a procedure of scheduling the charging process considering power limit is developed. The proposed method has been applied to and quantitatively evaluated in two application scenarios. Compared to randomly charging, both fluctuation and amplitude of the total current for multiple loads simultaneously being charged have been mitigated after properly scheduled. Using the proposed method, the merits of pulse charging for batteries can be utilized while the stability issue can be alleviated. 

\end{abstract}


\section*{Nomenclature} 
\subsection*{Variables} 
\addcontentsline{toc}{section}{Nomenclature}
\begin{IEEEdescription}[\IEEEusemathlabelsep\IEEEsetlabelwidth{$i, j, x, y$}] 
	\item[$D$] Duty ratio.
	\item[$f_n$] The frequency of pulse current $n$.
	\item[$I_{\rm \Sigma}$] Total charging current for multiple loads simultaneously being charged, ${\rm A}$.
	\item[$I_{\rm m}$] Amplitude of charging current through a load, ${\rm A}$.
	\item[$N_{\rm c}$] Number of loads simultaneously being charged.
	\item[$\overline{P}$] Mean power of the charging pulse in a period, ${\rm W}$.
	\item[$P_{\rm max}$] Upper limit for the total charging power, ${\rm W}$.
	\item[$P_{\rm \Sigma}$] Total charging power for all loads, ${\rm W}$.
	\item[$s$] The number of pulses or pulse currents regarded as ``bins'' or bin-type.
	\item[$t_{\rm on}$] Pulse width, ${\rm s}$.
	\item[$t_{\rm off}$] Length of the off-time interval in a pulse current, ${\rm s}$.
	\item[$T_i$] Period of pulse current $i$, ${\rm s}$.
	\item[$U_{\rm m}$] Amplitude of charging voltage on a load, ${\rm V}$.
	\item[$x_i$] A binary variable symbolizing whether an off-time interval is regarded as a ``bin'' or an ``item''.
	\item[$y_{ij}$] A binary variable indicating whether pulse $j$ can be turned-on during the off-time of pulse $i$.
	\item[$z_{ijk}$] A binary variable indicating if a pulse from sequence $j$ is turned-on during the $k^{\rm th}$ off-time interval of sequence $i$.
\end{IEEEdescription}

\subsection*{Abbreviations} 
\addcontentsline{toc}{section}{Nomenclature}
\begin{IEEEdescription}[\IEEEusemathlabelsep\IEEEsetlabelwidth{$i, j, x, y$}] 
	\item[EV] Electric vehicle.
	\item[FC] Fuel cell.
	\item[LCM] Least common multiple.
	\item[LIB] Lithium-ion battery.
	\item[PC] Pulse current.
	\item[PV] Photovoltaics.
	
\end{IEEEdescription}

\section{Introduction}
\IEEEPARstart{E}{lectric} vehicle (EV) sales have exceeded 17 million, accounting over 20\% of all cars sold worldwide in 2024, and over 40\% cars sold globally is predicted to be EV by 2030 \cite{IEA_GlobalEVOutlook2025}. As the EV market is booming, rapidly growing of the charging demand is expected. EVs consumes about 0.7\% of the world's total electricity, and this share is expected to reach 2.5\% by 2030 \cite{IEA_GlobalEVOutlook2025}. Charging of EVs not only increases the cumulative power level on electricity grid but also challenges the grid stability with fast transients \cite{Rivera2023}. Nevertheless, when properly coordinated, charging of EVs can also assist to support the grid through peak shaving, frequency regulation, voltage profile improvement, congestion management, etc. \cite{DEB2022}.

Conventional charging techniques for batteries include constant current (CC) charging, constant voltage charging (CV), and constant current-constant voltage (CC-CV)\cite{al-hajhussein2011,SHEN2012}. For Li-ion battery (LIB), the CC-CV technique is widely adopted thanks to its simplicity and easy implementation: the CC-CV technique keeps the charging current for LIB constant until a preset voltage of the LIBs is reached, then the charging voltage is hold constant while the charging current exponentially decreases, and the charging process ends when a minimum value of the current is reached\cite{SHEN2012}. Compared to the CC technique, the CC-CV technique allows LIBs to be almost fully charged and avoid overcharge, but it needs more time to finish the charging process and accelerates the capacity fading\cite{ABDELMONEM2015}. 

To boost up the charging velocity, various charging approaches have been proposed such as Multistage Constant Current (MCC) technique, which consists of several CC stages with different current levels. The MMC may limit the capacity fading and shorten the charging time through reducing generated heat and mechanical stress within LIBs during charging \cite{Tomaszewska2019}. Another fast charging technique called pulse charging, which offers lower rate of impedance rising compared to the MCC, has been discussed in literature. In this technique, pulse current (PC), which is periodically interrupted, is used instead of continuous current to charge the LIBs. The short pause set in each period offers a relaxation which can help eliminate concentration polarization, increase the power transfer rate, and thus reduce the charging time \cite{Li2001}. This technique has been claimed to be able to alleviate the capacity fading due to relaxation process set in this charging technique\cite{Lv2020}. The concentration is mitigated when the duration of pulse wave is in the range of seconds, because pulses shorter than milliseconds can be buffered by large double-layer capacitance at the electrode/electrolyte interface\cite{Chen2022}. Due to these merits, different charging systems using pulse charging have been developed such as presented in \cite{Chen2007,chen2008a,amanor-boadu2018}. 

However, the drastically varying and interrupted current and power of pulse charging may challenges the electric stability of the terminal during charging. This issue is critical to the power supply sources for which steady output power is in favor. For example, if photovoltaic (PV) panels are used to supply power for charging the LIBs, the frequently varying PC may disturb the operation point of the PV panels. In such case, the charging procedure needs to be carefully designed to deal with this inherent nature of interruption. To smoothly track the maximal power point of the PV panels, an circuit topology for pulse charging has been proposed in \cite{hsieh2014}, in which two additional batteries are explored to consume the positive PV power during the charging interruption and the reverse discharging power of the main battery. The interruption of PC can be suppressed or even eliminated when multiple batteries simultaneously being charged are properly coordinated. A control scheme regulating the buck converters controlling the pulse charging currents from fuel cell was proposed in \cite{jiang2004a}, in which several batteries are synergistically being charged to avoid drastic changing output power of fuel cell. Through successively charging two batteries using pulse charging, the magnitude and the variation of the total current used to charge the batteries can be reduced\cite{venkat2023}. To reduce the number of charger as well as its complexity, a circuit topology as well as a control scheme are proposed in \cite{liu2017}. By alternately charging the batteries, the number of chargers needed can be reduced, too. 

Proper scheduling the charging loads offers the potential to suppress the fluctuation and eliminate the intermittence of the charging current. However, the scheduling of multiple batteries using PC remains missing in the current literature. Aiming at this issue, a method for scheduling multiple batteries being charged at the same time is developed and presented in this paper with following contributions: First, a concept of adjusting the waveform of pulse charging current without changing the charging power is proposed. Second, two mathematical models and procedures for scheduling multiple loads simultaneously being charged using PCs with the same frequency and various frequencies are developed, respectively. Third, a procedure for coordinating charging loads considering power limit is proposed. 

This paper is organized as follows. Section \ref{Sec2} presents the concept of adjusting pulse current for a load without reducing the charging power. The idea of grouping and shifting the PCs for multiple loads to reduce the amplitude and fluctuation of the total current is presented in this section. After that, two scenarios regarding pulse frequency are considered. For each scenario, the grouping and shifting problem is mathematically modeled. Base on the mathematical solution, a procedure of scheduling the charging loads is developed. The proposed method is applied to two test simulations. The simulation results are presented and discussed in section \ref{Sec3}. Conclusions are drawn in section \ref{Sec4}.

\section{Optimized scheduling of pulse currents}\label{Sec2}
The periodic interruption of current introduced by the pulse charging offers batteries pauses to uniformly diffuse and distribute the electrolyte's ions\cite{liu2017}. However, this causes undesired intermittence of current and power supply at the charging terminal. When multiple batteries are being charged simultaneously, this is commonly the case for the batteries in a battery pack, such intermittence can be mitigated or even eliminated through properly coordinating the charging currents flowing to the batteries. In this section, an optimal coordination scheme is developed and elucidated in details. 

\subsection{The potential of staggering pulses from different currents}
\begin{figure}[htb]
	\centering
	\vspace{-0.5cm}
	\hspace{-0.5cm}
	\includegraphics[width=9cm]{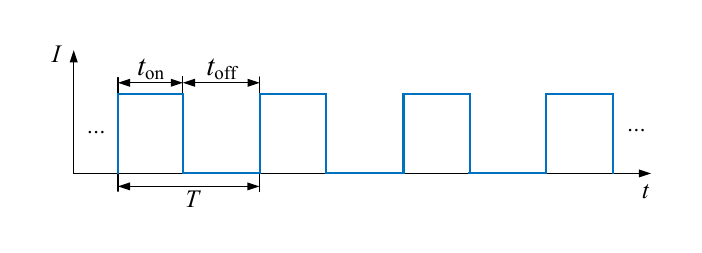}
	\vspace{-0.5cm}
	\caption{Example of the profile of a pulse current.}
	\label{PulseCurrent}
\end{figure}

The profile of a PC used in pulse charging can be approximately regarded as a sequence which consists of a series of rectangular pulses as shown in Fig. \ref{PulseCurrent}. In a PC, each pulse occurs at equal intervals within a period $T$ along the time axis, and each period covers a complete on-and-off cycle of a pulse. During one period, a pulse is active for a time interval of $t_{\rm on}$, after that the pulse is inactive for a time interval $t_{\rm off}$. In the on-stage, the charging current flows through a battery, and the battery will be charged. In the off-stage, the charging current falls to zero, and the charging process for the battery is temporally suspended. The ratio of $t_{\rm on}$ to $T$ is commonly defined as the duty ratio $D$ and mathematically expressed as:

\begin{equation}
	\label{DutyCycle}
	D = \frac{t_{\rm on}}{T} \times 100\%
\end{equation} 
	
A single pulse current introduces intermittence not only to the battery but also to the power supply source such as utility grid. This may pose a challenge to the stability of the power supply source, and this issue needs to be carefully dealt with. Nevertheless, when multiple batteries are charged using pulse charging at the same time, it is possible to adjust the PCs to eliminate this inherent intermittence. The idea is to shift the phases of each PC, so that the pulses from one PC can appear in the off-intervals of another PC. It should be noted that, in this paper, the term ``shift'' refers only to adjusting the initial phase of each PC and thus its relative position to another PC on the time axis. 

\begin{figure}[htb]
	\centering
	\vspace{-0.25cm}
	\includegraphics[width=9cm]{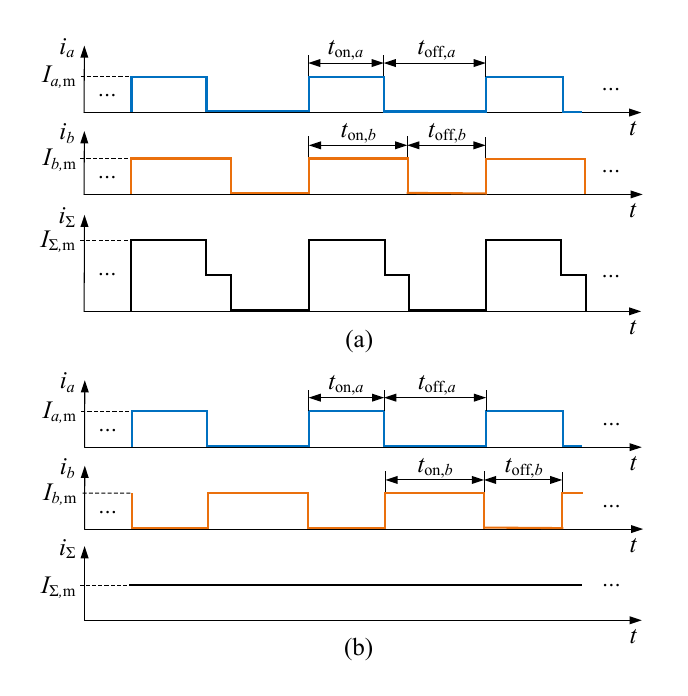}
	\vspace{-0.5cm}
	\caption{Adjusting the initial phases of two PCs: (a) Pulses aligned at the rising edge; (b) Pulses and off-intervals are complementary.}
	\label{Stagger2Profile}
\end{figure}

Taking two PCs at the same frequency whose profiles are shown in Fig. \ref{Stagger2Profile} as an example, there are different ways to adjust their initial phases in the time domain. One way is to set the pulses from PC $a$ and $b$ aligned to each other on the rising edge as depicted in Fig. \ref{Stagger2Profile} (a). In this case, the amplitude of the total charging current $i_{\rm \Sigma}$ is the sum of the amplitudes of both PCs, i.e. $I_{\rm \Sigma} = I_{a,{\rm m}} + I_{b,{\rm m}}$, where $I_{a,{\rm m}}$ and $I_{b,{\rm m}}$ are the amplitudes of PC $a$ and $b$, respectively. In another way, if the pulse widths of both PCs, $t_{\rm on,a}$ and $t_{\rm on,b}$, are not greater than the off-intervals of each other, i.e. $t_{\rm on,a} \leq t_{\rm off,b}$ and $t_{\rm on,b} \leq t_{\rm off,a}$, respectively, it is possible to stagger the pulses from both PCs on the time axis.

In the ideal case, when $t_{\rm on,a} = t_{\rm off,b}$, $t_{\rm on,b} = t_{\rm off,a}$, and $I_{a,{\rm m}} = I_{b,{\rm m}}$, the intermittence as well as the fluctuation of $I_{\rm \Sigma}$ can be eliminated by shifting the pulses from a PC to the off-intervals of another PC, as shown in Fig. \ref{Stagger2Profile} (b). In this case, $i_{\rm \Sigma}$ becomes a constant DC current and $I_{\rm \Sigma} = I_{a,{\rm m}} = I_{b,{\rm m}}$.    

\subsection{Adjusting the waveform of pulse current} \label{Adjustment}
\begin{figure}[htb]
	\centering
	\vspace{-0.5cm}
	\includegraphics[width=9cm]{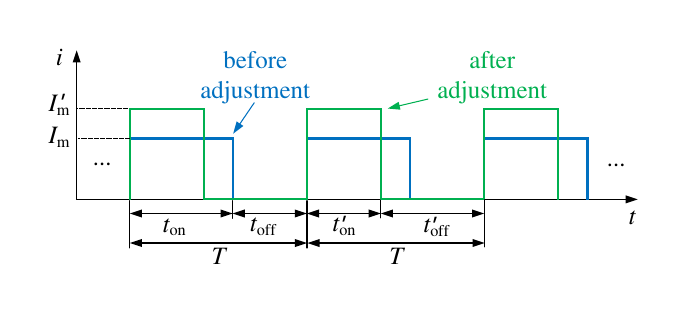}
	\vspace{-0.5cm}
	\caption{An example of waveform adjustment.}
	\label{WaveformAdjustment}
\end{figure}

As discussed above, it is possible to avoid the occurrence of pulses from two PCs at the same time by adjusting their initial phases. The condition, i.e. the pulse width of any PC is not greater than the off-interval of the other PC, must be satisfied. However, if this condition is not met, the waveforms of the PCs need to be adjusted. The adjustment can be implemented by tuning the period, the duty ratio, or/and the amplitude. To keep the frequency constant, only the duty ratio and the amplitude are changed here. An example of decreasing the duty ratio and increasing the amplitude without changing the period is shown in Fig. \ref{WaveformAdjustment}.

Before adjustment, the mean charging power within a period $\overline{I}$ can be calculated as:
\begin{equation}
	\overline{P} = \frac{t_{\rm on}}{T} U_{\rm m} I_{\rm m} = D U_{\rm m} I_{\rm m}
\end{equation}
where $U_{\rm m}$ and $I_{\rm m}$ are the amplitudes of the charging voltage and current, respectively, and $D$ is the duty ratio. As derived from the equation, $\overline{P}$ can be adjusted by tuning $D$, $U_{\rm m}$ or/and $I_{\rm m}$. After adjustment, the pulse width decreases from $t_{\rm on}$ to $t'_{\rm on}$, while the amplitude increases from $I_{\rm m}$ to $I'_{\rm m}$ with the new charging voltage $U'_{\rm m}$. The mean charging power becomes:
\begin{equation}
	\overline{P}' = \frac{t'_{\rm on}}{T} U'_{\rm m} I'_{\rm m} = D' U'_{\rm m} I'_{\rm m}
\end{equation}
where $D'$ is the duty ratio after the adjustment. This enables altering the charging power by adjusting the waveform of the PC. In addition, by changing the pulse width of the waveform, the condition required for staggering pulses from different PCs can be satisfied.

\subsection{Arranging pulse currents having the same frequency} \label{Stagger1}
\begin{figure}[htb]
	\vspace{-0.5cm}
	\includegraphics[width=9cm]{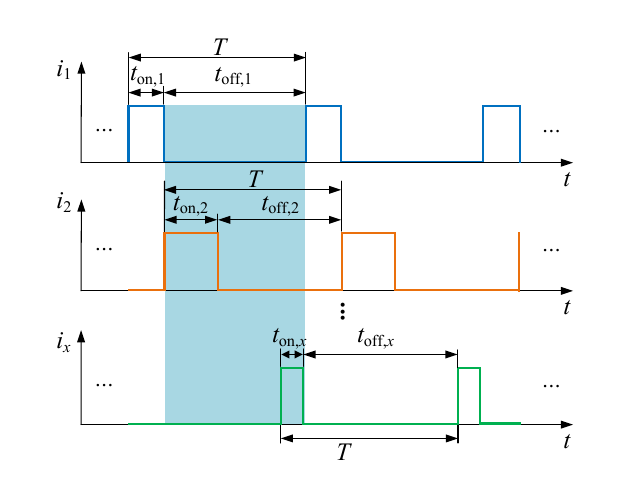}
	\vspace{-0.75cm}
	\caption{Scheme of staggering PCs at the same frequency.}
	\label{Pulses_SameT}
\end{figure}

The potential and precondition for shifting PCs to avoid the overlap of multiple pulses have been discussed in the previous section. Note that the PCs may have different frequencies. To begin with, this section focuses on the shifting of PCs at the same frequency as a simpler case. It is considered that $N_{\rm c}$ batteries are being charged using pulse charging at the same time. All PCs have the same frequency and amplitude, while the duty ratios may not necessarily identical to each other. Taking two PCs marked 1 and 2 as an example, if the pulse width of PC $2$ is not greater than the off-interval of PC 1, i.e. $t_{{\rm on,}2} \leq t_{{\rm off,}1}$, the pulse from PC $2$ can be shifted into the off-interval $t_{{\rm off,}1}$, as shown in Fig. \ref{Pulses_SameT}. Consequently, the overlap between the pulses from PC $1$ and PC $2$ is avoided. 

If the remaining part of $t_{{\rm off,}1}$ after subtracting $t_{{\rm on,}2}$ is not smaller than the pulse width from another PC $x$, i.e. $t_{{\rm off,}1} - t_{{\rm on,}2} \geq t_{{\rm on,}x}$, the pulse from PC $x$ can also be shifted into the interval $t_{{\rm off,}1}$ to stagger the pulses from PC 1 and 2, as shown in Fig. \ref{Pulses_SameT}. Generally, for a group of PCs at the same frequency, if the sum of the pulse widths from PC 2 to PC $x$ is not greater than the off-interval of PC 1, it is possible to shift the pulses from these $x-1$ PCs into the off-interval of PC 1 to avoid overlap between the pulses.

To mitigate the intermittence and fluctuation in the total charging current of multiple batteries, $i_{\rm \Sigma}$, the pulses from different PCs need to be uniformly distributed along the time axis. One way to realize this is to shift each pulse into the off-interval of another PC, so that the off-interval can be shortened or eliminated. In addition, the more pulses that can be shifted into the off-intervals from other PCs without overlapping, the more pulses can be prevented from turning on simultaneously, thereby reducing the peak value of $i_{\rm \Sigma}$. To optimally eliminate the intermittence and suppress the fluctuation of $i_{\rm \Sigma}$, a scheme for adjusting the phase the PCs needs to be developed. With such a scheme, the off-intervals can be filled with pulses from other PCs as much as possible. 

\begin{figure}[htb]
	\centering
	\vspace{-0.25cm}
	\includegraphics[width=9cm]{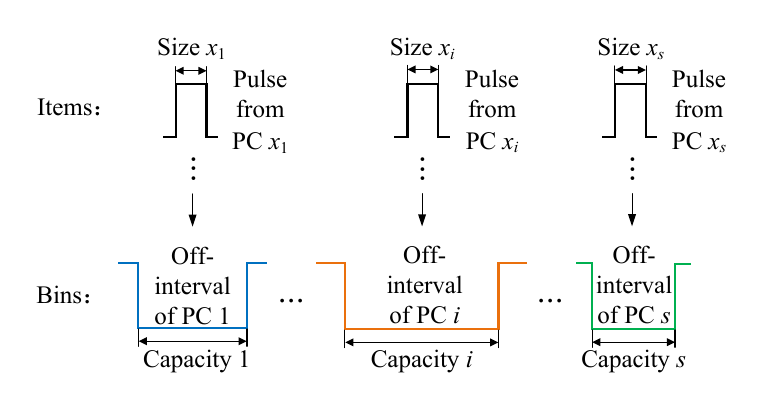}
	\vspace{-0.50cm}
	\caption{Analogy of shifting pulses at the same frequency to the classic bin packing problem.}
	\label{Analogy1}
\end{figure}

A pulse to be shifted can analogized as an item, while the off-interval can be regarded as a bin. Each item has a size determined by its pulse width, and each bin has a capacity corresponding to its time duration, as illustrated in Fig. \ref{Analogy1}. Accordingly, the problem of optimally shifting the pulses can be mapped to the classic bin packing problem, which aims to pack items of varying sizes into bins with fixed and predefined capacities, such that the number of bins used is minimized \cite{korte2005}. For simplicity reason, the PCs to be shifted are treated as items, and the off-intervals are treated as bins in this paper. To avoid double counting, each pulse is allowed to be assigned to at most one off-interval of another PC. 

Since all PCs operate at the same frequency and the pulses repeat within the same time period, it is sufficient to consider the problem for a single period. Thus, for each PC, only one representative pulse and its corresponding off-interval need to be considered. To mathematically formulate the grouping problem, several variables are introduced. The variable $x_i$ is used to indicate whether PC $i$ is treated as bin-type or item-type. When $x_i = 1$, PC $i$ is treated as bin-type, and pulses from other PCs can be shifted into its off-interval. Otherwise, $x_i = 0$ indicates that PC $i$ is treated as item-type, and the pulses from PC $i$ can be scheduled to turn on during the off-intervals of a bin-type PC. To describe the grouping relationship, an additional variable $y_{ij}$ is utilized to indicate whether the pulse from PC $j$ can be shifted to the off-interval of PC $i$. If the pulse from PC $j$ can be shifted to the off-interval of PC $i$ without overlapping with the pulse of PC $i$, $y_{ij} = 1$; otherwise $y_{ij} = 0$. 

For a real bin packing problem, if all items can be packed, it is obvious that fewer bins means less unused space in the bins. Determining a scheme to pack all items using the fewest bins is equivalent to minimizing the remaining space in the bins. Analogously, intermittence can be viewed as a form of unused space within the off-intervals. Therefore, to eliminate the intermittence, the off-intervals should be filled with pulses that can be shifted. If all pulses from the item-type PCs can be shifted to the off-intervals of the bin-type PCs, fewer bin-type PCs results in fewer unfilled off-intervals and thus less intermittence. Therefore, the optimal scheme for shifting the currents is consistent with minimizing the number of bin-type PCs. For $N_{\rm c}$ batteries, this objective can be mathematically expressed as:
\begin{equation}
	\label{obj1}
	{\rm min}\ s = \sum_{i=1}^{N_{\rm c}} x_i
\end{equation} 
subject to: \\
1.1: $\forall i \in [1,N_{\rm c}]$,
\begin{equation}
	\label{con1_1}
	\sum_{j=1}^{N_{\rm c}} \left(1-x_j\right) t_{{\rm on,}j} y_{ij} \leq x_i t_{{\rm off,}i},  
\end{equation} 
1.2: $\forall j \in [1,N_{\rm c}]$,
\begin{equation}
	\label{con1_2}
	\sum_{i=1}^{N_{\rm c}} y_{ij} = 1-x_j
\end{equation} 
where $s$ is the total number of bin-type PCs, $t_{{\rm on,}i}$ represents the pulse width of PC $i$, and $t_{{\rm off,}i}$ describes the widths of the off-interval of PC $i$. 

Constraint 1.1 is an inequality constraint that ensures the off-interval of PC $i$ is sufficiently wide to accommodate all pulses shifted to it without overlap. The left-hand side of the inequality represents the total width of all pulses that are shifted to the off-interval of PC $i$. Specifically, the pulse width of PC $j$ is included in this sum only if PC $j$ is classified as an item-type and it is shifted to the off-interval of PC $i$. In this case, $x_j = 0$, $y_{ij} = 1$. Therefore, $1 - x_j = 1$, and the pulse width $t_{{\rm on,}j}$ is added to the sum. The right-hand side of the inequality represents the width of the off-interval of PC $i$. It defines the capacity of the off-interval that treated as a ``bin'' to accommodate other pulses, which are treated as ``items'', without overlap. This capacity is considered only if PC $i$ is classified as a bin-type, which occurs when $x_i = 1$. 

Constraint 1.2 is an equality constraint. It ensures that each pulse can be shifted to at most one off-interval, thereby preventing overlaps or conflicting assignments.
The left-hand side of this equation sums up all variables $y_{ij}$. If PC $j$ is an item-type PC, $x_j = 0$, and the right-hand side of the equation equals 1. This ensures that any pulse from an item-type PC can only be shifted to exactly one specific off-interval of a bin-type PC. On the contrary, if PC $j$ is a bin-type PC, then $x_j = 1$, and the right-hand side of the equation becomes zero. As a result, all variables $y_{ij}$ for any $i$ associated with PC $j$ must be 0. This guarantees that no pulse from a bin-type PC is allowed to be shifted to the off-interval of any other PC. In other words, the bin-type PCs only provide their off-intervals for accommodating pulses from the item-type PCs, but do not contribute their own pulses for relocation.

By solving the objective function \cref{obj1} and the constraints \cref{con1_1} - \cref{con1_2}, a grouping scheme of $N_{\rm c}$ PCs operating at the same frequency can be determined. First, for each bin-type PC $i$ with $x_i = 1$, all item-type PCs $j$ with $y_{ij} = 1$ are selected. These PCs are grouped together as group $i$, where PC $i$ is the bin-type PC and the other PCs are the item-type PCs in the group. Then, the pulses from the item-type PCs are shifted to the off-interval of the bin-type PC in the group and activated successively without overlapping, as shown in Fig. \ref{Pulses_SameT}.

\subsection{Arranging PCs having various frequencies} \label{Stagger2}
In some cases, different batteries may be charged using PCs operating at different frequencies to minimize the battery impedance, since batteries have different internal resistance \cite{Chen2007}. The duty ratios of different PCs are not equal to each other and may also vary with time \cite{Chen2009}. In this section, PCs operating at different frequencies are considered. To mitigate the intermittency and fluctuation in the total charging current for multiple batteries, which are being charged using PCs at various frequencies, a grouping and pulse-shifting scheme is developed and explained in this section. 

\begin{figure}[htb]
	\centering
	\vspace{-0.5cm}
	\includegraphics[width=9cm]{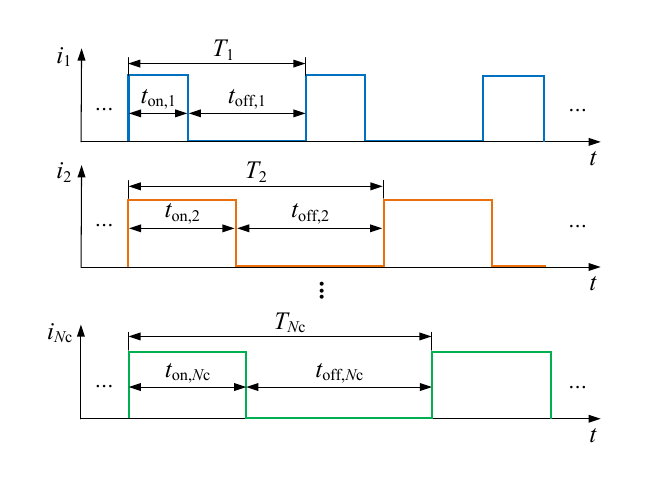}
	\vspace{-0.5cm}
	\caption{Profile of PCs at different frequencies.}
	\label{Pulses_DifferentT}
\end{figure}

A set of PCs operating at different frequencies and duty ratios is illustrated by the profiles in Fig. \ref{Pulses_DifferentT}. Since the pulses from different PCs may occur at different times, the relative positions of pulses and off-intervals from different PCs on the time axis are not fixed. Therefore, shifting the pulses from a PC to the off-intervals of another PC on the time axis is feasible only if the following two conditions are satisfied. First, there exists at least one PC whose pulse width is less than or equal to the off-interval of at least one other PC in the group. This condition is straightforward: if the width of every pulse is greater than all off-intervals in the group, then it is impossible to shift any pulse into an off-interval without causing overlap of pulses. Second, for any two PCs that satisfy the first condition, their periods must be such that the larger period is an integer multiple of the smaller one. This condition ensures that if a pulse from PC $a$ can be shifted to an off-interval of PC $b$, then all pulses from PC $a$ can also be shifted to the off-intervals of PC $b$.  

\begin{figure}[htb]
	\centering
	\hspace{-0.5cm}
	\includegraphics[width=9.25cm]{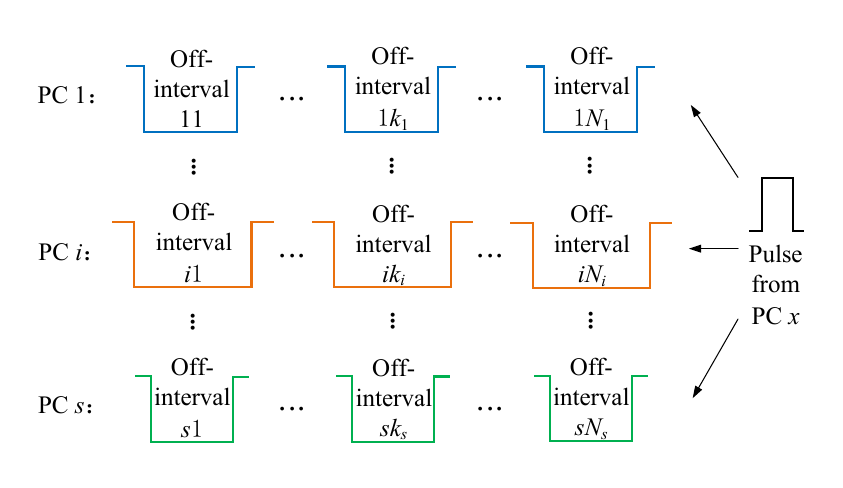}
	\caption{Analogy of the staggering of pulses at different frequencies to the bin packing problem.}
	\label{Analogy2}
\end{figure}

Once both conditions are satisfied, all pulses in the group can be scheduled to avoid overlap. Similarly, the grouping and shifting problem can be analogized to the bin packing problem. However, in this case, the PCs operate at different frequencies. Therefore, it is no longer sufficient to analyze the problem within a single period of each PC. From the perspective of all PCs collectively, their patterns repeat every $T_{\rm LCM}$, which is the least common multiple (LCM) of their individual periods. During $T_{\rm LCM}$, each off-interval of a PC can be treated as a bin and has the potential accommodate pulses from other PCs. The pulses shifted to the off-interval can be analogized as items packed into a bin, as illustrated in Fig. \ref{Analogy2}. Note that, since the PCs have different frequencies, the numbers of off-intervals for each PC within $T_{\rm LCM}$ varies, i.e. the number of ``bins'' differs across PCs. 

Similar to the previous case, to eliminate the intermittence inherent in each PC, the off-intervals of each PC are preferably be filled with pulses from other PCs as much as possible. Analogous to the bin-packing problem, where the goal is to minimize the number of bins and the unused space by packing all items efficiently. Furthermore, to mitigate the fluctuation resulting from the sum of all PCs, their positions in time need to be properly adjusted to avoid overlap wherever possible. To achieve these two goals, a strategy for grouping the PCs and scheduling their phases must be designed. 

To formulate a mathematical model for the problem, several variables are introduced. As in the previous case, the variable $x_i$ indicates whether PC $i$ is a bin-type or an item-type PC. When $x_i = 0$, PC $i$ is classified as an item-type PC, and its pulses need to be shifted to the off-intervals of other PCs. Otherwise, $x_i = 1$ indicates that PC $i$ is a bin-type PC, and its off-intervals can accommodate pulses from item-type PCs. Similarly, variable $y_{ij}$ is to used to indicate whether pulses from PC $j$ can be shifted to the off-intervals of PC $i$. If the pulses from PC $j$ can be shifted to the off-intervals of PC $i$ without overlapping the pulses of PC $i$, then $y_{ij} = 1$; otherwise, $y_{ij} = 0$. Additionally, a variable $z_{ijk}$ is used to indicate whether a pulse from PC $j$ can be shifted to the $k^{\rm th}$ off-interval of PC $i$. Specifically, if this is feasible, then $z_{ijk} = 1$; otherwise $z_{ijk} = 0$. 

To minimize the total charging current for $N_{\rm c}$ batteries that are charged at the same time using pulse charging method, the objective can be mathematically expressed as:
\begin{equation}
	\label{obj2}
	{\rm min}\ s = \sum_{i=1}^{N_{\rm c}} x_i
\end{equation} 
subject to: \\
2.1: $\forall k \in [1,N_i]\ {\rm and}\ \forall i \in [1,N_{\rm c}]$,
\begin{equation}
	\label{con2_1}
	\sum_{j=1}^{N_{\rm c}} \left(1-x_j\right) t_{{\rm on,}j} z_{ijk} \leq x_i t_{{\rm off,}i}  
\end{equation} 
where $t_{{\rm off,}i}$ is the duration of the off-interval of PC $i$; \\
2.2: $\forall n \in [1,N_i]\ {\rm and}\ \forall i,j \in [1,N_{\rm c}]\ {\rm with}\ T_j = R_{ji} T_i$, and $R_{ji}$ is positive integer,
\begin{equation}
	\label{con2_2}
	\sum_{k=n+1}^{n+R_{ji}} z_{ijk} = y_{ij}
\end{equation} 
where $T_i$, $T_j$ are the periods of PC $i$ and $j$, respectively. \\
2.3: $\forall j \in [1,N_{\rm c}]$,
\begin{equation}
	\label{con2_3}
	\sum_{i=1}^{N_{\rm c}} y_{ij} = \left(1-x_j\right)
\end{equation} 

The inequality constraint 2.1 guarantees that any off-interval in PC $i$ is long enough to accommodate all pulses shifted to it without overlap. The left-hand side of inequality constraint (\ref{con2_1}) is the sum of the width of all pulses shifted to the $k^{\rm th}$ off-interval of PC $i$. The width of the pulses from PC $j$ is included in the sum when the PC is NOT bin-type and it is assigned to PC $i$, i.e. $x_j = 0$ and $y_{ij} = 1$. In addition, this pulse must be shifted to the $k^{\rm th}$ off-interval in PC $i$, i.e. $z_{ijk} = 1$. Only in this case $\left(1-x_j\right) z_{ijk} = 1$ and $t_{{\rm on,}j}$ is added to the sum. The right-hand side of the inequation represents the width of the off-interval of PC $i$. The width is taken into account only when PC $i$ is considered as bin-type, and thus $x_i = 1$. It is to be emphasized that this inequity constraint must be valid for ALL off-intervals included in the period of $T_{\rm LCM}$. 

The equality constraint 2.2 ensures that the period of PC $j$ is equal to $T_j$. The left-hand side of \cref{con2_2} sums up the values of any $R_{ji}$ adjacent $z_{ijk}$, whose subscripts $k$ are consecutive. This sum equals to the number of pulses from PC $j$ shifted to any $R_{ji}$ consecutive off-intervals in PC $i$. Since $T_j = R_{ji} T_i$, there is only one off-interval within a period $T_j$ in PC $j$, while $R_{ji}$ off-intervals are included in the same period ot time in PC $i$. This means that any pulse from PC $j$ can be shifted to only one of every $R_{ji}$ consecutive off-intervals. Therefore, the sum of any $R_{ji}$ adjacent $z_{ijk}$ must be equal to 1. If PC $j$ is not grouped with PC $i$, then $y_{ij} = 0$ and thus the sum of $z_{ijk}$ is always 0. This is consistent with the fact that no pulse of PC $j$ is shifted to any off-interval in PC $i$.

The equality constraint 2.3 ensures that PC $j$ is grouped with only one other current, if PC $j$ is item-type. The left-hand side of \cref{con2_2} is the sum of all $y_{ij}$ indicating whether PC $j$ is grouped with PC $i$. To avoid duplicate shifting of a pulse to multiple off-intervals, this sum is limited to not greater than 1. The right-hand side of the equality constraint can either be value 1 or 0 depending on whether PC $j$ is item-type or bin-type, respectively. 

\begin{figure}[htb]
	\centering
	\hspace{-0.5cm}
	\includegraphics[width=9.25cm]{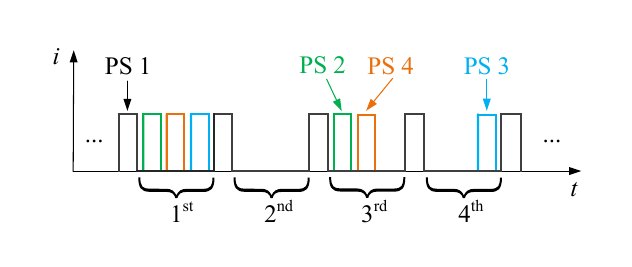}
	\vspace{-0.5cm}
	\caption{An example of grouping and shifting PCs that have different frequencies.}
	\label{Example1}
\end{figure}

By solving \cref{obj2} - \cref{con2_3}, the grouping and shifting scheme for the PCs having different frequencies can be determined. First, for each certain value of $i$, if $x_i = 1$, then PC $i$ and all PCs $j$ with $y_{ij} = 1$ are picked out and grouped together. Then, for each PC in the same group, the value of $z_{ijk}$ indicates the relative location of PC $j$ to PC $i$. Specifically speaking, the pulse from PC $j$ need to be shifted to the $k^{\rm th}$ off-intervals of PC $i$ with $z_{ijk} = 1$. For example, $x_1 = 1$ and $y_{12} = y_{13} = y_{14} = 1$ suggest that PCs 1,2,3, and 4 need to be put into the same group. In this case, if $z_{121} = z_{123} = 1$ and $z_{122} = z_{124} = 0$, then the pulses from current 2 needs to be shifted to the 1$^{\rm st}$ and 3$^{\rm rd}$ off-intervals in current 1, and no pulse from current 2 is shifted to the 2$^{\rm nd}$ and 4$^{\rm th}$ off-intervals, as shown in Fig. \ref{Example1}. Similarly, with $z_{131} = z_{134} = z_{141} = z_{143} = 1$, $z_{132} = z_{133} = z_{142} = z_{144} = 0$, the shifting scheme for pulse current 3 and 4 is as depicted in Fig. \ref{Example1}. 

The shifting scheme shown in Fig. \ref{Example1} can be implemented through adjusting the initial phases of the PCs in the same group, so that their relative position can be uniformly distributed to avoid overlap of pulses. If the initial phase of PC $i$ is $\phi_{0,i}$, its $k^{\rm th}$ off-interval begins at $t = \phi_{0,i} + t_{{\rm on},i} + (k-1) \times T_i$, this can be set as the initial phase $\phi_{0,j}$ of PC $j$. Generally, if the first pulse of PC $j$ is shifted to the $k^{\rm th}$ off-interval of PC $i$ just after the $m^{\rm th}$ pulse in the off-interval, then the initial phase of PC $j$ can be set as $\phi_{0,j} = \phi_{0,i} + t_{{\rm on},i} + (k-1) \times T_i + \sum_{l=1}^{m}t_{{\rm on},l}$, which can be set as the initial phase of this PC. The initial phase of other PCs can be similarly determined.

\subsection{Grouping PCs at different frequencies}
\begin{figure}[htb]
	\centering
	\hspace{-0.5cm}
	\includegraphics[width=8.5cm]{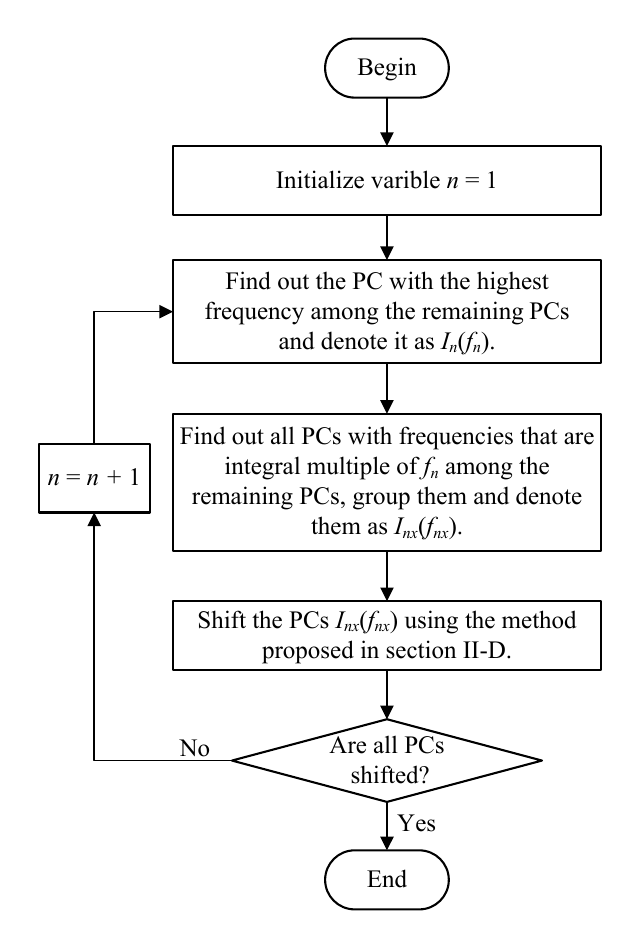}
	\vspace{-0.5cm}
	\caption{The procedure for arranging PCs that have different frequencies.}
	\label{Procedure1}
\end{figure}

As discussed before, when the PCs have different frequencies, they can not necessarily be grouped together and shifted without overlap of their pulses. The stagger of the PCs is feasible when both two conditions mentioned in the previous section are fulfilled. Therefore, before shifting the PCs, they need to be divided into different groups first. To do this, following procedure is proposed:
\begin{enumerate}
	\item initialize an integer variable $n$ as $n = 1$;
	\item pick out the PC which has the highest frequency from the remaining PCs and denote it as $p_n(f_n)$, where $f_n$ is the frequency of PC $n$;
	\item find out from the remaining PCs all PCs at frequencies $f_{nx}$ which are integral multiple of $f_n$, where $f_{nx}$ refers to the frequencies of the pulses and $x$ is a consecutive positive integer for each PC. Group the PCs being found out and denote them as $p_{nx}(f_{nx})$;
	\item shift the PCs $p_{nx}(f_{nx})$ using the method proposed in section \ref{Stagger2};
	\item if there are PCs remaining not grouped, make $n = n + 1$ and move to step 2), repeat the procedure until all PCs have been grouped and shifted. 
	The whole procedure is visually shown as the flow chart in Fig. \ref{Procedure1}.
\end{enumerate}

\vspace{-0.5cm}

\subsection{Coordinating charging loads considering power limit} \label{Priority}
\begin{figure}[htb]
	\centering
	\vspace{-0.5cm}
	\includegraphics[width=9cm]{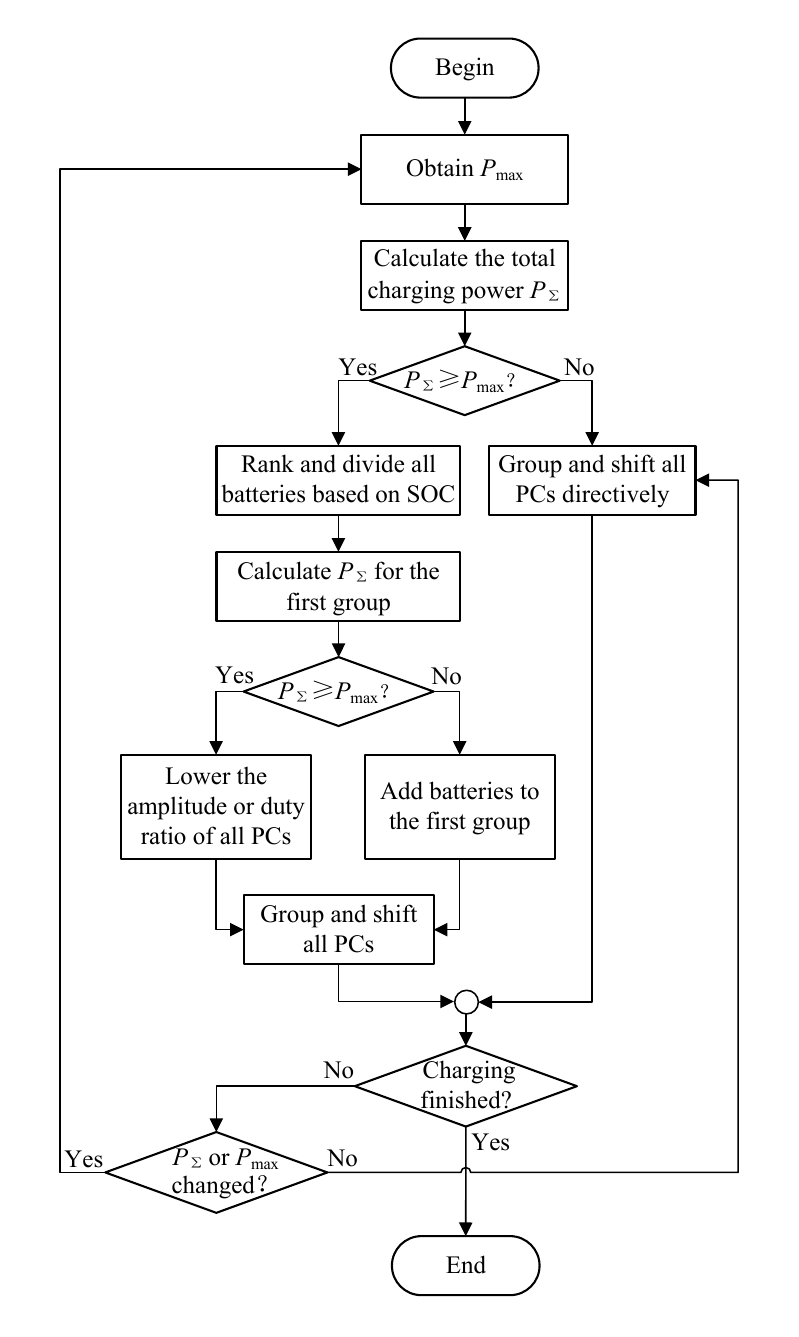}
	\vspace{-0.5cm}
	\caption{The procedure for coordinating loads considering power limit.}
	\label{Procedure}
\end{figure}

In some cases, the charging power for batteries needs to be regulated to support coordinate with the power supply source or electricity grid by engaging in demand side management. In this section, a procedure for coordinating a series of batteries being charged is proposed. During charging, the total power $P_{\Sigma}$ needed for all batteries is calculated by summing up the charging power for each battery. If $P_{\Sigma}$ is smaller than $P_{\rm max}$, which is the upper power limit for $P_{\Sigma}$. If $P_{\Sigma}$ exceeds $P_{\rm max}$, $P_{\Sigma}$ needs to be reduced. To reduce $P_{\Sigma}$, the batteries can be divided into two groups according to their state-of-charge (SOC). The batteries in the first group have lower SOC than that in the second group. Therefore, charging for the batteries in the first can be initiated prior to the batteries in the second group. The total charging power $P_{\Sigma}$ for the first group is calculated and monitored in real-time during charging. Once $P_{\Sigma}$ excesses $P_{\rm max}$, $P_{\Sigma}$ can be further reduced by lowering the amplitude $I_{{\rm m},i}$ or duty ratio $D_{i}$ of each PC $i$ to $I^{-}_{{\rm m},i}$, $D^{-}_{i}$, respectively, as follows:
\begin{equation}
	\label{I_reduced}
	I^{-}_{{\rm m},i} = \frac{P_{\rm max}}{P_{\Sigma}} I_{{\rm m},i}
\end{equation} 
\begin{equation}
	\label{D_reduced}
	D^{-}_{i} = \frac{P_{\rm max}}{P_{\Sigma}} D_{i}
\end{equation} 
After that, $P_{\Sigma}$ is reduced down to $P_{\rm max}$. On the other hand, if $P_{\Sigma}$ is below $P_{\rm max}$, some batteries from the second group can be added to the first group. The number $m$ of the batteries added to the first group needs to obey the following inequity constraint:
\begin{equation}
	P_{\rm max} - P_{\Sigma} \le \sum_{i=1}^{m} P_i
\end{equation} 
The coordination procedure is visually depicted in the flow chart shown in Fig. \ref{Procedure}.

The charging for the second group is postponed until either $P_{\rm max}$ is increased or the charging of any battery in the first group is finished. Once the priority of charging is determined, the grouping and shifting PC for each battery can be performed using the procedures proposed in previous sections. If all batteries are charged using PCs at the same frequency, the grouping and shifting scheme for the PCs can be determined using the procedure proposed in section \ref{Adjustment}. If the PCs have different frequencies, they can be grouped and shifted using the procedures proposed in section \ref{Stagger2} and this section. After the scheme is determined, the charging can be started.

\section{Application}\label{Sec3}
In this section, the proposed method for scheduling of charging loads is applied in two test scenarios. In the first scenario, all loads are charged using PCs at the same frequency, while in the second scenario the PCs having various frequencies. In both scenarios the PCs are grouped and staggered to alleviate the overlap of pulses using the proposed methods. 

\subsection{Scenario 1: all PCs have the same frequency}
\begin{figure}[htb]
	\centering
	\vspace{-0.5cm}
	\includegraphics[width=9.5cm]{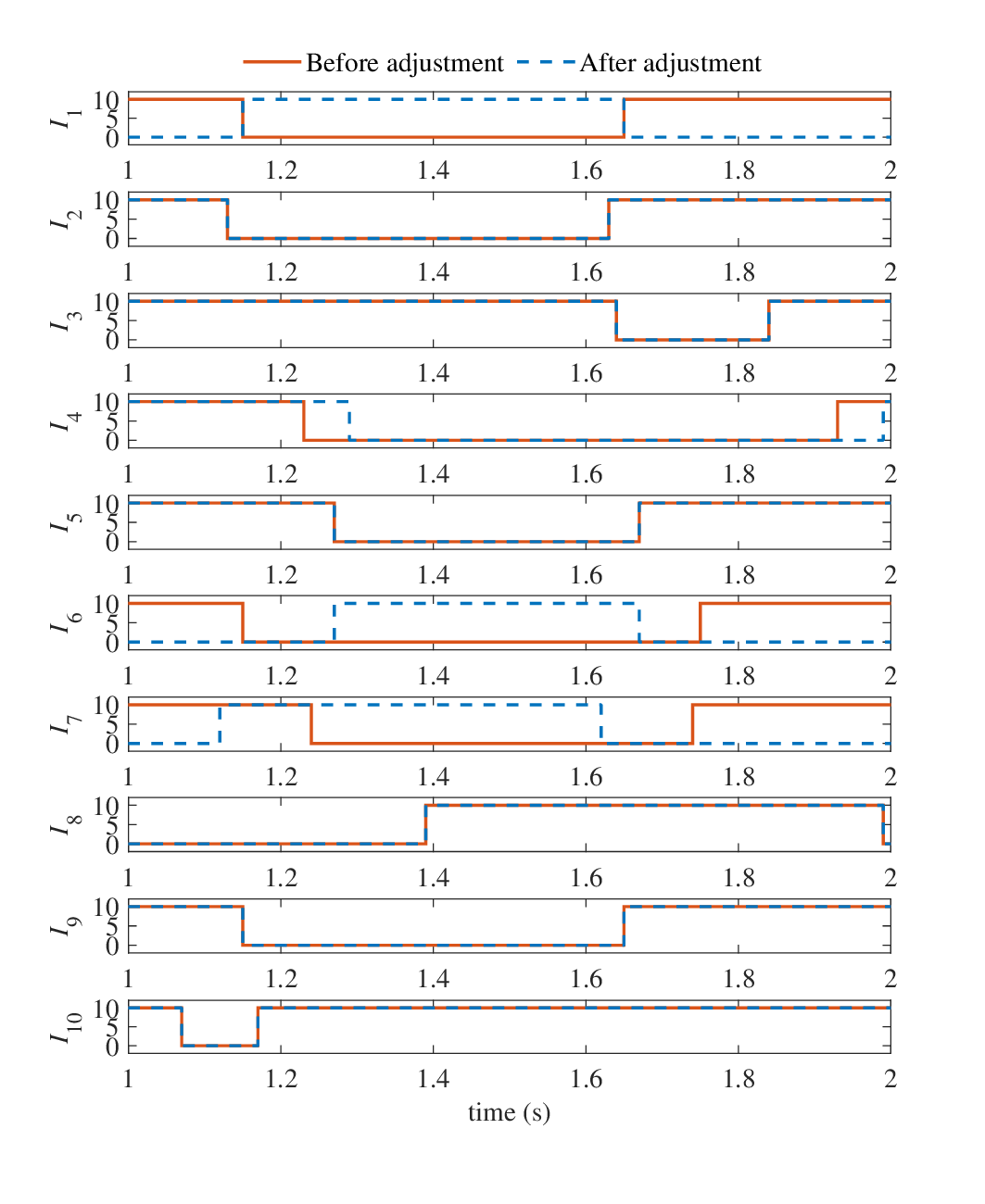}
	\vspace{-1.25cm}
	\caption{Profiles of PCs at the same frequency before and after the staggering.}
	\label{I_i1}
\end{figure}

In the first scenario, ten loads to be charged using PCs at the same frequency are considered. The parameters related to the PCs used to charge each load are listed in Table \ref{Parameter1}. The profiles of the PCs are plotted by the orange solid lines shown in Fig. \ref{I_i1}.
\begin{table*}[ht]
	\caption{Parameters of the PCs in scenario 1.} 
	\label{Parameter1} 
	\centering
	\begin{tabular}{|p{5cm}<{\centering}|p{0.5cm}<{\centering}|p{0.5cm}<{\centering}|p{0.5cm}<{\centering}|p{0.5cm}<{\centering}|p{0.5cm}<{\centering}|p{0.5cm}<{\centering}|p{0.5cm}<{\centering}|p{0.5cm}<{\centering}|p{0.5cm}<{\centering}|p{0.5cm}<{\centering}|}
		\hline \hline
		Pulse & 1 & 2 & 3 & 4 & 5 & 6 & 7 & 8 & 9 & 10 \\
		\hline \hline
		Amplitude (A) & \multicolumn{10}{c|}{10} \\
		\hline
		Frequency (Hz) & \multicolumn{10}{c|}{1} \\
		\hline
		Duty ratio (\%) & 50 & 50 & 80 & 30 & 60 & 40 & 50 & 60 & 50 & 90 \\
		\hline
		Random initial phase (s) & 0.65 & 0.63 & 0.83 & 0.93 & 0.67 & 0.75 & 0.74 & 0.39 & 0.65 & 0.17 \\
		\hline
		Initial phase after staggering (s) & 0.15 & 0.63 & 0.84 & 0.99 & 0.67 & 0.27 & 0.13 & 0.39 & 0.65 & 0.17 \\
		\hline \hline
	\end{tabular}
\end{table*}

\begin{figure}[htb]
	\centering
	\includegraphics[width=9cm]{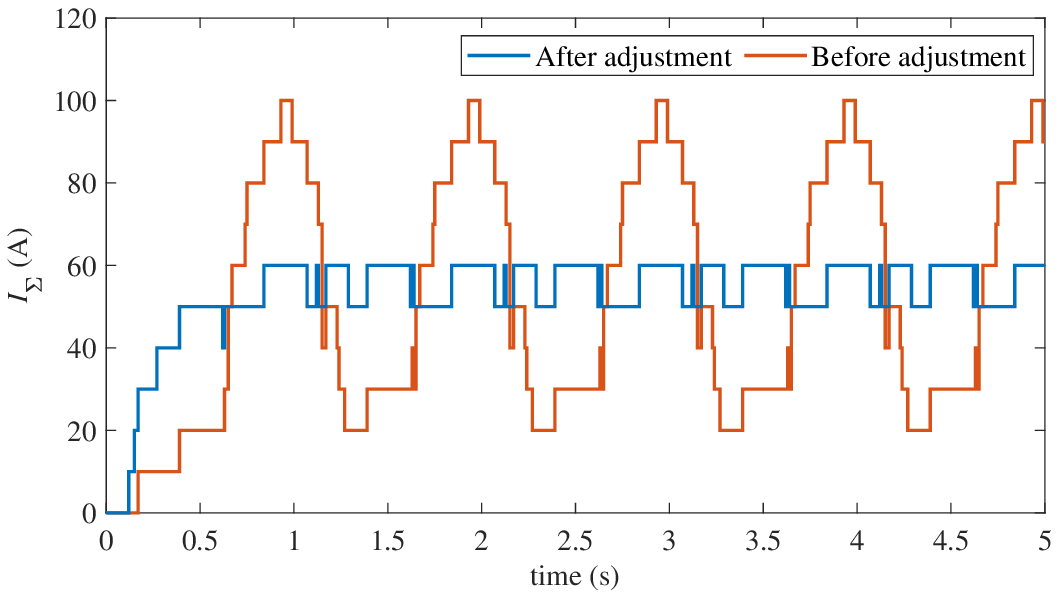}
	\caption{The procedure for coordinating loads considering power limit.}
	\label{I_sum1}
\end{figure}

To begin with, the charging of all loads are unordered, i.e. all PCs are located on the time axis with random initial phases as given in the table. The total charging current of the ten loads ranges from 20 A to 100 A, as plotted in the solid line in orange shown in Fig. \ref{I_sum1}. 

Noted that all PCs have the same frequency, to reduce the fluctuation of the total charging current $I_{\Sigma}$, the grouping and staggering method presented in section \ref{Stagger1} is utilized to adjust the relative position of all PCs. After the adjustment, the initial phase of the PCs are obtained and given in Table \ref{Parameter1}. A comparison of the profiles of each pulse current are visually presented in Fig. \ref{I_i1}. As can be seen from the figure, the profiles of pulses 1, 4, 6, and 7 have been shifted along the time axis. 

The total charging current for the ten loads after staggering is as plotted in the blue solid line shown in Fig. \ref{I_sum1}. As can be seen, the maximum value of $I_{\Sigma}$ is reduced from 100 A to 60 A, while the minimum value of $I_{\Sigma}$ is increased from 20 A to 50 A, the fluctuation of $I_{\Sigma}$ is mitigated. The total charging current for the ten loads after staggering is as plotted in the blue solid line shown in Fig. \ref{I_sum1}. As can be seen, the maximum value of $I_{\Sigma}$ is reduced from 100 A to 60 A, while the minimum value of $I_{\Sigma}$ is increased from 20 A to 50 A, the fluctuation of $I_{\Sigma}$ is mitigated.

\subsection{Scenario 2: PCs having various frequencies}
\begin{figure}[htb]
	\centering
	\vspace{-0.5cm}
	\includegraphics[width=9.5cm]{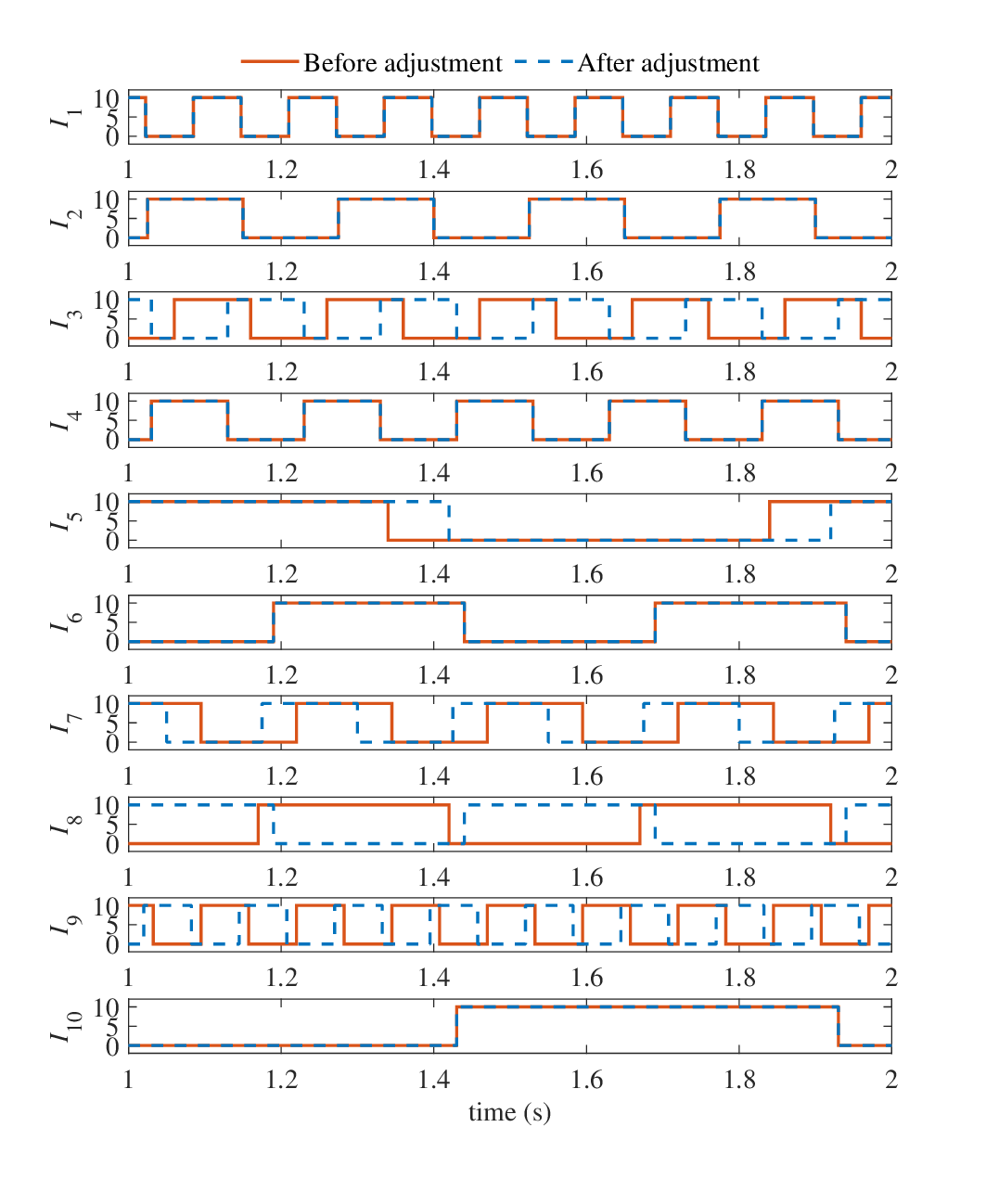}
	\vspace{-1.25cm}
	\caption{Profiles of PCs having different frequencies before and after the staggering.}
	\label{I_i2}
\end{figure}

Similar to the first scenario, ten loads to be charged using PCs are investigated. Differs to the first scenario, however, the frequencies of the PCs are not all the same. The parameters of the PCs are listed in Table \ref{Parameter2}. The profiles of the PCs are plotted by the orange solid lines shown in Fig. \ref{I_i2}.
\begin{table*}[ht]
	\caption{Parameters of the PCs in scenario 2.} 
	\label{Parameter2} 
	\centering
	\begin{tabular}{|p{5cm}<{\centering}|p{0.5cm}<{\centering}|p{0.5cm}<{\centering}|p{0.5cm}<{\centering}|p{0.5cm}<{\centering}|p{0.5cm}<{\centering}|p{0.5cm}<{\centering}|p{0.5cm}<{\centering}|p{0.5cm}<{\centering}|p{0.5cm}<{\centering}|p{0.5cm}<{\centering}|}
		\hline \hline
		Pulse & 1 & 2 & 3 & 4 & 5 & 6 & 7 & 8 & 9 & 10 \\
		\hline \hline
		Amplitude (A) & \multicolumn{10}{c|}{10} \\
		\hline
		Frequency (Hz) & 8 & 4 & 5 & 5 & 1 & 2 & 4 & 2 & 8 & 1 \\
		\hline
		Duty ratio (\%) & 50 & 50 & 50 & 50 & 50 & 50 & 50 & 50 & 50 & 50 \\
		\hline
		Random initial phase (s) & 0.21 & 0.30 & 0.47 & 0.23 & 0.84 & 0.19 & 0.22 & 0.17 & 0.22 & 0.43 \\
		\hline
		Initial phase after staggering (s) & 0.21 & 0.30 & 0.33 & 0.23 & 0.93 & 0.19 & 0.42 & 0.44 & 0.27 & 0.43 \\
		\hline \hline
	\end{tabular}
\end{table*}

\begin{figure}[htb]
	\centering
	\includegraphics[width=9cm]{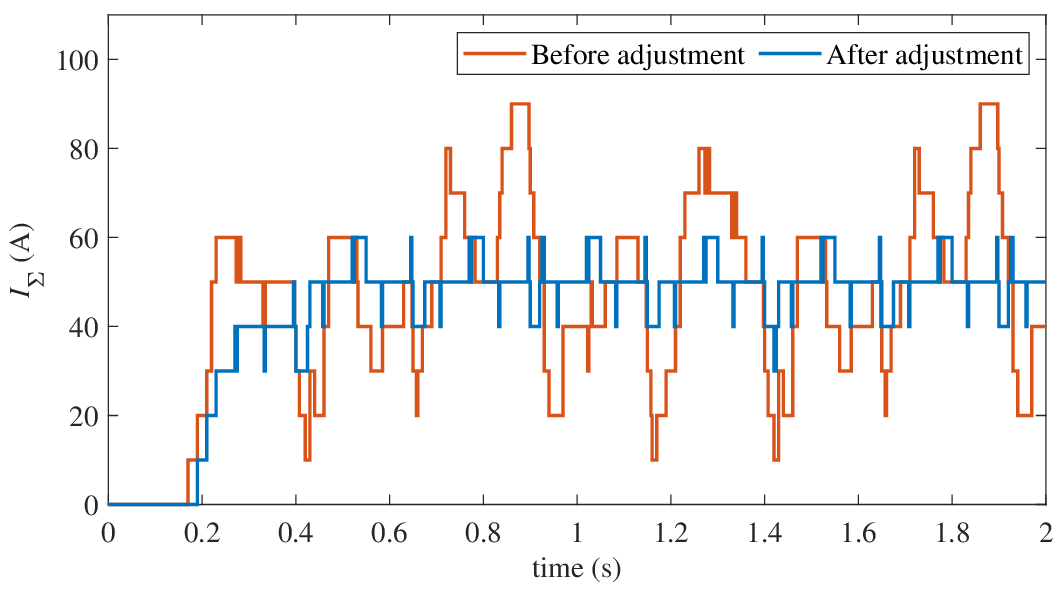}
	\caption{The procedure for coordinating loads considering power limit.}
	\label{I_sum2}
\end{figure}

Based on Table \ref{Parameter2}, it can be calculated that the periods of the PCs are 1 s, 0.5 s, 0.25 s, 0.2 s, and 0.125 s. It is not difficult to obtain the least common multiple of all the periods $T_{\rm LCM} = 1$ s. Initially, the PCs are randomly flowing through the loads resulting a profile of the total current $I_{\Sigma}$ of all loads as depicted by the orange solid line shown in \ref{I_sum2}. As can be seen, in this case, $I_{\Sigma}$ fluctuates from 10 A to 90 A.  
Though not all PCs have the same frequency, however, some pulses have met the two conditions presented in section \ref{Stagger2}. Therefore, it is possible to stagger these pulses to alleviate the overlap of pulses. The grouping and staggering method proposed in section \ref{Stagger2} can thus be utilized to adjust the relative position of all pulses, so that the fluctuation of the total charging current $I_{\Sigma}$ can be reduced. To realize this, the initial phase of the PCs need to be adjusted as given in Table \ref{Parameter2}. The profiles of each pulse current before and after adjustment are plotted in Fig. \ref{I_i2}. As can be seen from the figure, the phases of pulses 3,5,7,8 and 9 have been shifted along the time axis, while the phases of the other pulses remain unchanged. 

After the shifting, the minimum and maximum values of $I_{\Sigma}$ have been changed from previous 10 A and 90 A to 40 A and 60 A, respectively. The profile of $I_{\Sigma}$ is depicted by the blue line shown in Fig. \ref{I_sum2}. It is clearly demonstrated by the two test scenarios that both the peak values and the fluctuation of $I_{\Sigma}$ can be reduced using the proposed method.

\section{Conclusions}\label{Sec4}
To mitigate the fluctuation and intermittence of current for multiple loads simultaneously being charged using pulse charging method, a method for coordinating the charging process for the loads has been proposed in this paper. First, the concept of adjusting pulse current without changing the charging power was introduced. Then the potential of exploiting the off-time intervals of pulse current to charge other loads was explained. Through properly grouping the loads and shifting the charging currents on the time axis, the current pulses can be staggered. Thereafter, to find out the optimal charging scheme for all loads, the scheduling problem was mathematically described. Two scenarios were considered and two correspondent mathematical models have been proposed. In one scenario all loads are charged using PCs with the same frequency, and in the other scenario PCs with various frequencies are used. In addition, a procedure of scheduling the charging process considering power limit is developed. The proposed method has been applied to and quantitatively evaluated in two application scenarios. Compared to randomly charging, both fluctuation and amplitude of the total current for multiple loads simultaneously being charged have been mitigated after properly scheduled. Using the proposed method, the merits of pulse charging for batteries can be utilized while the challenge to stability of electric power supply can be addressed.

%
%
%
%
%

\ifCLASSOPTIONcaptionsoff
  \newpage
\fi

\bibliography{IEEEabrv,Reference}
\begin{IEEEbiography}{Yu Liu}
	received the B.Sc. degree from Xi'an Jiaotong University in electrical engineering, Xi'an, China, in 2010. He received the M.Sc. and Dr.-Ing. degrees in electrical engineering from Technische Universit\"at Berlin, Berlin, Germany, in 2015 and 2020, respectively. He is currently an Assistant Professor with Shenzhen Technology University, Shenzhen, China. His research interests include integration of renewable energy sources into electric networks, modeling and simulation of power systems, energy interaction between different energy systems, analysis of digital simulation system.
\end{IEEEbiography}

\end{document}